\documentclass[aps, showpacs,twoside,floatfix,a4,superscriptaddress]{revtex4}
\usepackage{bbm}
\usepackage{amsmath, amssymb}
\usepackage{color}
\usepackage{graphicx,epsfig}
\usepackage{pifont} 

\newcommand{\nmqc}{{NMQC$_{\oplus}$}}
\newcommand{\nmqcs}{{NMQC$_{\oplus}^*$}}
\newtheorem{lem}{Lemma}

\newcommand{\ket}[1]{|#1\rangle}

\begin{document}

\title{Non-adaptive Measurement-based Quantum Computation and Multi-party Bell Inequalities}

\author{Matty J. Hoban}
\email{m.hoban@ucl.ac.uk}
\affiliation{Department of Physics and Astronomy, University College London, Gower Street, London WC1E 6BT, United Kingdom.}
\author{Earl T. Campbell}
\affiliation{Department of Physics and Astronomy, University College London, Gower Street, London WC1E 6BT, United Kingdom.}
\affiliation{Institute of Physics and Astronomy, University of Potsdam, 14476 Potsdam, Germany.}
 \author{Klearchos Loukopoulos}
\affiliation{Department of Materials, Oxford University, Parks Road, Oxford OX1 4PH, United Kingdom.}
\author{Dan E. Browne}
\affiliation{Department of Physics and Astronomy, University College London, Gower Street, London WC1E 6BT, United Kingdom.}
\affiliation{Centre for Quantum Technologies, National University of Singapore, Singapore, 117543.}

\date{\today}

\begin{abstract}
\noindent
Quantum correlations exhibit behaviour that cannot be resolved with a local hidden variable picture of the world. In quantum information, they are also used as resources for information processing tasks, such as Measurement-based Quantum Computation (MQC). In MQC, universal quantum computation can be achieved via adaptive measurements on a suitable entangled resource state. In this paper,  we look at a version of MQC in which we remove the adaptivity of measurements and aim to understand what computational abilities still remain in the resource. We show that there are explicit connections between this model of computation and the question of non-classicality in quantum correlations. We demonstrate this by focussing on deterministic computation of Boolean functions, in which natural generalisations of the Greenberger-Horne-Zeilinger (GHZ) paradox emerge;  we then explore probabilistic computation, via which multipartite Bell Inequalities can be defined. We use this correspondence to define families of multi-party Bell inequalities, which we show to have a number of interesting contrasting properties.
\end{abstract}



\maketitle

\noindent

\section{Introduction}

\noindent
Space-like separated measurements on entangled quantum  systems can reveal correlations which would not appear if Nature were described by a local hidden variable (LHV) theory \cite{bell}. Such correlations can  violate  a Bell inequality, a condition expressing a restriction on the correlations permitted in any LHV theory. Violations of Bell inequalities provide a dramatic distinction between locally realistic classical physics and quantum mechanics.

The field of quantum information processing has developed with the aim of exploiting the non-classical properties of quantum systems for practical tasks. {Non-classical} (i.e. Bell inequality violating) correlations are an important quantum resource in several areas of quantum information. For example, they can be used to guarantee the security of  quantum key distribution \cite{ekert} even if the devices used cannot be trusted; in distributed computational tasks, non-classical correlations can reduce the communication required between parties \cite{brukner, buhrman}. One of the most renowned applications of quantum mechanics to information processing is quantum computation. We know that a quantum computer, if realised, would permit efficient solutions to problems for which no efficient classical algorithm is known \cite{shor}. Despite the importance of entanglement in quantum computation, and despite its evident non-classical nature, the link between Bell inequality violations and quantum computational speed-up remains unclear. Recently, however, it was shown \cite{anders,raussendorf} that Bell inequalities do have a concrete connection with a particular model of quantum computation, measurement-based quantum computation (MQC). 

MQC is a model of quantum computation, very different from the standard unitary circuit model. Computation progresses in two stages. First a many-particle entangled state is prepared, e.g. the cluster state \cite{rausbrowne}, and then single-site (qubit) measurements are performed. By an appropriate choice of measurement bases, any quantum computation can be realised in this model on a cluster state of sufficient size \cite{rausbrowne}. 

In all schemes of MQC that have been proposed so far (e.g. \cite{briegel,grosseisert}) measurement bases need to be \emph{adaptive}. Adaptive measurements are made in a certain order, and each measurement basis is chosen dependent on the outcome of the measurements which came before it. A vital component of MQC schemes, which is sometimes overlooked, is the classical computer, which records measurement outcomes, and determines \cite{rausbrowne} the bases of future measurements. The adaptivity imposes a temporal structure on the computation, and adds the requirement for fast-switching of measurement bases in any experimental implementation. It is therefore worthwhile considering whether MQC without adaptive measurements is possible, and if it is, what is the range (and complexity) of computations it can achieve. This would present significant practical advantages since the quantum coherence and entanglement would only need to exist for a vanishingly small time, since all measurements could be made simultaneously.  Recently, some progress was made in characterising the power of non-adaptive MQC, and it was shown \cite{browneperdrickashefi} that one can implement efficiently, any computation in the complexity class IQP (a model known as instantaneous quantum computation) \cite{shepherd}. There is evidence \cite{shepherd2} that IQP may contain algorithms with a quantum speed-up, however, no concrete example is yet known.

An interesting fact about the adaptivity in Raussendorf and Briegel's MQC scheme \cite{briegel} is that the side-processing needed to implement the adaptivity does not require a universal classical computer. It suffices for the computations to be implemented solely by {exclusive OR} (XOR) gates. In other words, such computations are linear Boolean functions \cite{anders}. This motivated Anders and Browne to consider what the minimal (in the sense of the number of qubits) quantum correlations capable of enabling universal classical computation with a linear side processor. Surprisingly they discovered that the correlations required are those which define the both Clauser-Horne-Shimony-Holt (CHSH) inequality \cite{chsh}, and  Greenberger-Horne-Zeilinger (GHZ) paradox \cite{ghz}. 

There are a number of superficial similarities between experiments that aim to produce a Bell inequality violation, and MQC. In both cases, the initialisation of the experiment consists of the generation of an entangled state; the experiment then proceeds by the  measurement of individual subsystems (usually qubits) in one of a range of bases. However, there is also an important distinction. In Bell inequality experiments, measurements must be space-like separated, whereas in MQC adaptivity precludes this. In this paper, therefore, we shall restrict our our focus to non-adaptive MQC, where this conflict does not arise.

\begin{figure}
	\centering
		\includegraphics[width=0.50\textwidth]{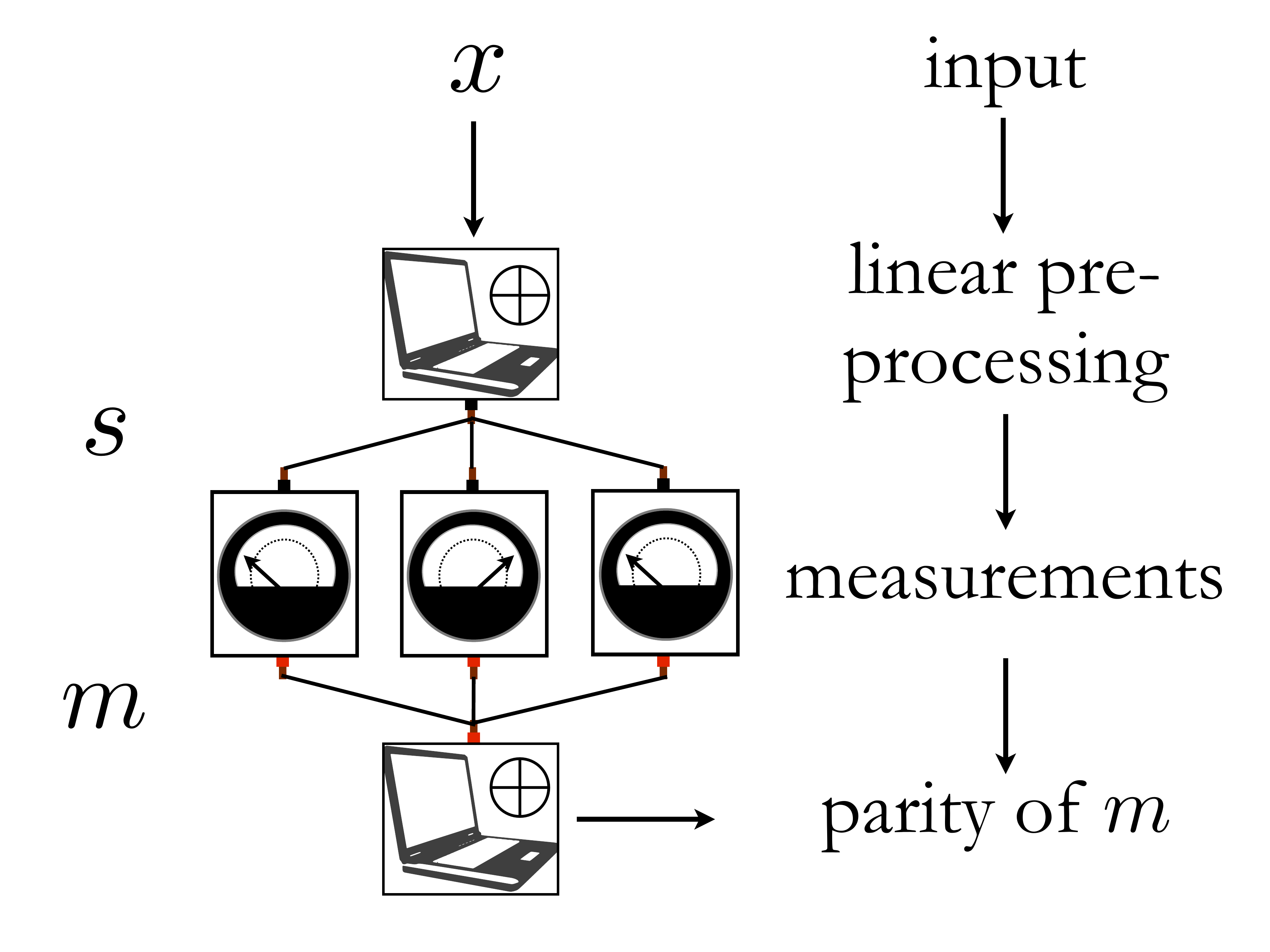}
			\caption{A schematic representing the operations which make up an {{\nmqc}} computation. The input is a bit-string $x$. The classical side processor then sends to each $j^{\textrm{th}}$ measurement device, a bit $s_{j}$, corresponding to the output of linear function on $x$. It receives measurement outcomes (as bits) from each device and then, after some linear post-processing returns the output of the computation.}
	\label{fig:belltest}
\end{figure}

Since non-adaptive MQC with linear side-processing will play a central role in this paper, it is worthwhile introducing a shorthand to refer to it. We therefore define the model of \textit{Non-adaptive Measurement-based Quantum computation with linear side-processing} ({{\nmqc}}). This is the standard MQC model with two key restrictions: measurements are non-adaptive, and the classical side processor responsible for pre- and post-processing of data is limited to  linear functions alone. The model of {{\nmqc}} is represented schematically in Figure. \ref{fig:belltest}. In this paper, we will study the range of computations which can be achieved within {{\nmqc}} both deterministically, and non-deterministically, and study the Bell inequalities that these computational tasks define. The importance of restricting the side-computation to linear functions is that we can study
 classical computations (which we will always describe in terms of Boolean functions) in the {{\nmqc}} model.

It has already been shown, by Raussendorf \cite{raussendorf}, that general non-linear Boolean functions may be achieved in a model equivalent to {{\nmqc}}, and each of these can be seen as a particular multi-qubit generalisation of a GHZ paradox. 
GHZ paradoxes can be thought of as representing special cases of Bell inequality violations in the many-party setting. We show here that a broad class of natural multi-party generalisations of the CHSH inequality, where measurements are restricted to two settings and two outcomes, can be cast (as computational non-locality ``games'') directly in the {{\nmqc}} framework.
Doing this immediately provides new structures for the classification of Bell inequalities, and allows us to derive general properties of $n$-party families of Bell inequalities true for all $n$.


Although the connection between MQC and Bell inequalities is recently made, there is a long tradition of viewing Bell inequalities as information theoretic tasks, for example, in the context of quantum games \cite{cleve, annereview}. This approach is to frame the Bell inequality experiment in terms of the achievement of a certain task by a set of players, who are given certain resources and who must abide by certain rules. In Bell inequality games, the players represent the physicists making the measurements. The rules of the game include a requirement that these players do not communicate during the task (or be space-like separated) and the resource might be an entangled quantum state shared between players. In the context of Bell inequalities, we compare two kinds of resource available to the players. Players with access to only classical resources represent the behaviour found in local hidden variable theories, whereas players with access to entangled states represent quantum behaviours. When a quantum player beats a classical one in a certain game (usually by achieving a higher mean success probability in a repeated task), this represents the violation of a Bell inequality. The quantum games approach to Bell inequalities has been very successful, and, in particular, the family of XOR games, generalisations of the CHSH inequality,  where the players encode the output of their task in the parity of the bits that they produce, are well characterised and understood \cite{cleve}.

Here we show that {{\nmqc}}  naturally defines a family of multi-party quantum games, which, in turn can be seen to be equivalent to a broad family of generalised CHSH inequalities. The task in the game will be the completion of a computation (in the {{\nmqc}} model) specified by a particular Boolean function $f(x)$ on an $n$-bit string $x$. In section \ref{det} we study tasks, which, with quantum resources can be performed deterministically under {{\nmqc}}. Furthermore, we show that all functions can be achieved in this model, and characterise an upper bound on the necessary number of measurements required for a generic function which is not always efficient in the size of $x$. Then we shall focus on two specific families of functions, and calculate lower bounds on the number of measurements which can be achieved. This setting, in which tasks, which succeed deterministically with quantum resources, but which are impossible to achieve deterministically with LHV or classical correlations, are natural generalisations of the GHZ ``all-or-nothing'' refutation of LHV models.

In section \ref{nondet}, we shall consider non-deterministic games. If we attempt to compute a function $f(x)$ in {{\nmqc}} with less than the minimal number of measurements required for a deterministic computation, then we know that the computation can only succeed with less than unit probability. However, we can still calculate the best mean success probability of the game for quantum and classical resources. Bell inequalities can be associated with the upper bounds on these games for classical correlations. We shall show that the {{\nmqc}} framework captures all of the broad family of generalisations of the CHSH introduced by Werner, Wolf \cite{WW}, \.{Z}uchowski and Brukner \cite{ZB}. We shall then use it to calculate upper bounds on success with quantum resources, or potential \emph{quantum violation} of the classical, LHV upper bound for the families of functions identified in the previous sections, including the surprising result that the (arguably) most natural instance of a generalisation of the CHSH inequality to the many-party setting has no quantum violation for $n>2$. We conclude, in section \ref{discussion} with a discussion of the the consequences of results and open questions which remain.

We shall adopt the following notation throughout this paper. For bit-string $x$, $x_j$ will represent the $j$th bit of $x$ and $|x|$ will represent its length. $W(x)$ will represent the Hamming weight (number of non-zero bits) of $x$.

\section{The  {{\nmqc}} model}

We begin by formalizing {{\nmqc}}, as outlined in Fig.~\ref{fig:belltest}.  The goal is to compute a Boolean function $f(x)$ given an $n$-bit input bit-string $x$. 
  The available resource is a quantum state on which we may perform spatially separated measurements.  Each measurement can be performed in one of two bases, e.g. $M_{0}$ or $M_{1}$, and so we denote the choice of all measurement bases and their measurement outcomes by bit strings $s$ and $m$ respectively where the elements of each are each $j^{\textrm{th}}$ party's choice of measurement $s_{j}$ and outcome $m_{j}$. The length of these bit-strings $|s|=|m|$ is determined by the number of measurements. For a non-adaptive model, the measurement settings cannot depend on outcomes of other measurement outcomes, but they  depend only on the input string $x$.  All side-processing must be linear, in this model, so the measurement settings, $s$, depend only linearly on $x$.  Without loss of generality we can assume the linear function is homogenous, and write
\begin{equation}
	s = ( P x )_{\oplus}
\end{equation}
where $P$ is an $n$-by-$|s|$ binary matrix and addition is modulo 2.  Any  {{\nmqc}} protocol is completely described by this matrix $P$, the quantum state used and the  measurement bases.  Note that is essential that the measurement settings, $s$, have a non-trivial dependence on the input, $x$, since a fixed basis measurement would be merely equivalent to a  classical device outputting bits according to a classical probability distribution. Finally, the output of the computation is achieved by then   linear post-processing on values of $m$, typically the parity of $m$. Deterministic {{\nmqc}} is achieved when the parity of $m$ always gives $f(x)$ for all values $x$.

\section{Deterministic computations and GHZ paradoxes}\label{det}

\noindent
Recall that every Boolean function $f(x)$ on an $n$-bit string $x$ can be represented uniquely as a polynomial over the finite field $Z_2$ (i.e. using modulo 2 arithmetic). This polynomial is known as the algebraic normal form (ANF) for the function. The ANF allows one to naturally classify Boolean functions according to the degree of the associated polynomial. Linear functions are functions of degree 1, i.e. of the form $f(x)=a_0+\bigoplus_{j=1}^n a_j x_j$, where $a_k$ are bits and $x_j$ is the $j^{\textrm{th}}$ bit of $x$. A non-linear function is one which cannot be written in this form. 

It is not a priori clear that this very restricted model of {{\nmqc}} would be particularly expressive. After all, it seems few unitary circuits in MQC can be achieved deterministically without adaptive measurements. Nevertheless, in \cite{raussendorf}, Raussendorf showed, within a framework essentially identical to {{\nmqc}}, that non-linear functions can be computed by measurements on Reed-Muller states, a certain family of stabilizer states, and furthermore, that each non-linear function led to a GHZ-like paradox, i.e. its deterministic evolution would be impossible to achieve with LHV correlations. 

Our first result is to show that, in fact, \emph{all} Boolean functions $f(x)$ can be achieved deterministically in {{\nmqc}}. Furthermore, our result is constructive, identifying the resource states, and a means to compute the measurement bases to compute the function. It also provides an upper-bound to the number of measurements needed to implement any Boolean function deterministically in {{\nmqc}}. 
\newline

\noindent
\textbf{Theorem 1} In {{\nmqc}} with quantum resources, any Boolean function $f(x)$, on an arbitrary $n$-bit string $x$, can be computed with unit probability. This can be achieved with the resource state $\frac{1}{\sqrt{2}}(\ket{0}^{\otimes (2^{n}-1)}+\ket{1}^{\otimes (2^{n}-1)})$, ie. a $(2^{n}-1)$ qubit generalised GHZ state. 
\newline

\noindent
\textit{Proof of Theorem 1} Our proof is constructive. Recall that in {{\nmqc}} each measurement device (or party) receives a bit-value $s_j$, which is a linear function on input bit-string $x$. Each then makes their measurement accordingly, returning the output of their measurement $m_{j}$ (as a bit-value) to the side processor.
Here, the $(2^{n}-1)$ parties share a generalised GHZ state, as defined above.
The $j^{\textrm{th}}$ party will receive bit value $s_j$. They will then make a measurement in the following basis $\cos(s_j\phi_j)\sigma_{x}+\sin(s_j\phi_{j})\sigma_{y}$, where $\phi_j$ is an angle which is yet to be specified. Note that a 0 input, in all cases, corresponds to a measurement of $\sigma_x$. They will report the output of these measurements as a bit-value (e.g. 0 for eigenvalue $+1$, and 1 for eigenvalue $-1$), which is returned to the side processor.

The output of the {{\nmqc}} computation $f(x)$ will be encoded (deterministically) in the parity of these outcome bits from the joint outcome of all parties. A simple analysis of the properties of GHZ states (see also \cite{WW}) shows us that the parity of the output bits will deterministically be equal to $f(x)$ if the following equation is satisfied
\begin{equation}\label{theorem1key}
e^{i\sum_j s_j(x)\phi_j}=
{-1}^{f(x)+c}, \forall x \in \{ 0, 1 \}^{n}
\end{equation}
where $s=( P x )_{\oplus}$. The bit $c\in\{0,1\}$ represents a linear operation that may be implemented in post-processing, and we can thus, without loss of generality restrict $f(x)$ to functions where $f(0^{\otimes n})=0$ and add any additional bit-flip in post-processing.  The exponent on the L.H.S  of \eqref{theorem1key} is a sum over real numbers (in the angles $\phi_{j}$) whereas the term in the exponent of the right-hand side is a Boolean function, which we are expressing as a polynomial over $Z_{2}$, i.e. with addition modulo 2. We shall exploit the fact that these different types of addition lead to different notions on linear independence. 
To emphasize this distinction we refer to linear (in-)dependence over the reals as $\mathbb{R}$-linear (in-)dependence and linear (in-)dependence over $Z_2$ as $Z_2$-linear (in-)dependence.

The set of linear Boolean functions on $x$, which appear in the sum on the left hand side of the equation,  are not necessarily $Z_2$ linearly independent, but, as we shall show below, all such functions are linearly independent over the reals. This means that linear functions (over $Z_2$) can be combined linearly (in the real vector space) to produce a \emph{non-linear} Boolean function. It is this mathematical fact which allows non-linear functions to appear on the right hand-side of equation \eqref{theorem1key} and forms the basis for the ability for non-linear functions to be computed within {{\nmqc}}.

Since this effect may seem at first hand surprising, we provide an explicit example. Consider the AND function $x_1x_2$ on two bits $x_{1}$ and $x_{2}$. This can be written as a linear combination of $Z_2$-linear functions as follows
\begin{equation}\label{paritytoreal}
	x_1 x_2 = \frac{1}{2}\left(x_1 + x_2 - (x_1\oplus x_2)
	\right).
\end{equation}

This is an instance of a more general identity expressing the parity of a bit-string in real arithmetic
\begin{equation}\label{modtoreal}
 	\bigoplus_i x_i=\frac{1}{2}[1-\prod_i(1-2x_i)]=\sum_{b}(-2)^{W(b)-1}\prod_{j}^{n}x_{j}^{b_{j}} ,
\end{equation}
where we sum over all $n$-length bit-strings $b$ and $W(b)$ is the Hamming weight of $b$.  We complete our proof by showing that within the linear Boolean functions on bit-string $x$ we can find a complete basis (over the reals) for the vector space of real-valued functions on $x$ and thus any Boolean function $f(x)$ can be represented as a $\mathbb{R}$-linear combination of them.

An example of such a basis is the set of functions $f_a(x)=\bigoplus_{j=1}^n a_jx_j$ for all bit strings $a$ (except the all zeroes string), augmented by the constant function $f_0(x)=1$. We see that this set has the required number of elements $2^n$ to form a basis, so it merely remains to prove that these are $\mathbb{R}$-linearly independent. To do so, we exploit the fact that the monomial functions (i.e. Boolean functions with a single multiplicative term) are linearly independent in both $\mathbb{R}$ and $Z_{2}$. 

Every monomial function can be expressed as a real linear combination of parity functions. To see this, notice that
equation \eqref{modtoreal} shows us that  every function $f_a(x)$, when expressed in arithmetic over the reals, is a $\mathbb{R}$-linear combination of a single monomial $\prod_{j\in a,a_{j}=1} x_j$ of degree $W({a})$ and monomials of degree less than $W({a})$. 
Thus the degree $W({a})$ monomial is a linear combination of $f_a(x)$ and monomials of lesser degree. Since the degree 2 monomials can by equation \eqref{paritytoreal} be expressed as linear combinations of parity functions, inductive reasoning tells us that all higher degree monomials must also expressible as real linear combinations of parity functions alone. Since the monomials are linearly independent with respect to the reals, this implies that the functions $f_a(x)$ for all $n$-bit strings $a$ do indeed form a basis for $n$-bit functions over the reals. Furthermore, this implies that all functions where $f(0^{\otimes n})=0$  can be achieved by $\mathbb{R}$-linear combinations of parity functions $f_a(x)=\bigoplus_{j=1}^n a_jx_j$.

Hence solutions of equation \eqref{theorem1key} exist for any Boolean function $f(x)$ on $n$ bits. Solving \eqref{theorem1key} provides the measurement angles $\phi_j$ needed to implement the computation. This completes the proof.$\square$
\newline

We see that the generalised GHZ states are universal resources for Boolean functions, in a similar sense to the way cluster states are universal resources for quantum computation.  GHZ states always form the optimal resource for computations in {\nmqc}, in the sense that they minimise the number of qubits required for the computation. This is very closely related to Werner and Wolf's proof that GHZ states always provide maximal violations of any generalised CHSH inequality \cite{WW} as shall be touched upon later in this paper.
\newline

\noindent
\textbf{Corollary 1}: Raussendorf shows in \cite{raussendorf} that non-entangled quantum states can only facilitiate deterministic {{\nmqc}} computations for degree 1 functions. Raussendorf shows in \cite{raussendorf} that  non-entangled quantum states can  only facilitate deterministic {{\nmqc}} computations for degree 1 functions. Thus the achievement of a non-linear Boolean function in {{\nmqc}} requires entanglement and may be considered (as discussed in \cite{raussendorf}) a type of GHZ paradox.
\newline

Theorem 1 provides an upper-bound to the number of qubits required to perform any Boolean function in {{\nmqc}}. Unfortunately, this bound is exponential in the size $n$ of the input string, and this computation strategy is therefore {inefficient}. It is pertinent to ask whether this many qubits are necessary, or whether, for some functions, a more compact measurement scheme can be found. We shall now therefore calculate, for two families of functions, the lower bounds in the number of measurements required for deterministic evaluation under {{\nmqc}}.

The families we consider are both natural multipartite generalisations of the two-bit AND function:

\begin{itemize}
	\item \textbf{$n$-tuple AND function:} $g_{n}(x)=\prod_{j=1}^{n}x_{j}$. This function computes the joint product of all bits.
\item \textbf{Pairwise AND function:} $h_{n}(x)=\bigoplus_{j=1}^{n-1}x_{j}(\bigoplus_{k=j+1}^{n}x_{k})$. This function computes the sum (modulo 2) of the pairwise product of all pairs of bits.
\end{itemize}

Studying these families of functions, we are able to prove the following two lower bounds:
\newline

\noindent \textbf{Proposition 1:} To achieve the $n$-tuple AND function $g_{n}(x)$ deterministically in {{\nmqc}} one requires no fewer than $(2^{n}-1)$ qubits.
\newline

We prove this result in two steps, provided in Appendices \ref{app_P_ Equivalence} and \ref{app1}.  First we demonstrate (Appendix  \ref{app_P_ Equivalence} ) that many different choices of linear preprocessing, as described by the $P$ matrix, are equivalent.  Furthermore, that when $|s|<(2^{n}-1)$ the $P$ matrix is always equivalent to a canonical form with a specific property.  Next we show that  (Appendix  \ref{app1}), given this canonical form of $P$, satisfying all $2^{n}$ required equations would entail a contradiction, and hence, must be impossible.
\newline

\noindent\textbf{Proposition 2:} To achieve the pairwise AND function $h_{n}(x)$ deterministically in {{\nmqc}}  one requires no fewer than $(n+1)$ qubits.
\newline

\textit{Proof}: The resource which achieves this bound is an $(n+1)$ qubit GHZ state $\frac{1}{\sqrt{2}}(\ket{0}^{\otimes n}\otimes\ket{1}+\ket{1}^{\otimes n}\otimes\ket{0})$ where the qubits are numbered $j=1,2,...,n,n+1$ from left-to-right. A simple calculation will verify that with input bits consisting of $s_{j}=x_j$ for $j=1,\ldots,n$ and $s_{n+1}=\bigoplus_j x_j$ for the $(n+1)$th site and measurement bases $\sigma_x$ or $\sigma_y$ for inputs 0 or 1 respectively, the parity of all $(n+1)$ outputs is always equal to $h_n(x)$.

We now prove that this is a lower bound. We do so by showing that no $n$-bit non-linear function may be achieved deterministically in \nmqc with $n$ measurements alone.
Given $n$ independent two-setting measurements on an $n$-subsystem state, there are $2^n$ possible combinations of measurements. If the parity of all measurement outcomes for all these $n$ measurement sites is to be deterministic for every combination of settings, the state must be an eigenstate of every different $2^n$ tensor product combination of the local observables, with eigenvalue $(-1)^{f(x)}$. But, since these are dichotomic measurements, the local observables $M_{j}$ are idempotent i.e. $M_{j}^{2}=\mathbb{I}$. Hence by considering all products of the operators (i.e. finding the full stabilizer of the state), one quickly finds that the joint eigenstate must also be an eigenstate of local unitaries on each site, i.e. must be a product state. Since we know that unentangled states are not resources in {{\nmqc}} for non-linear functions, $(n+1)$ measurements must be a lower bound on any deterministic non-linear function in {{\nmqc}}. $\square$
\newline

\begin{figure}
\begin{tabular}{ l l } a) & b) \\ \includegraphics[width=0.40\textwidth]{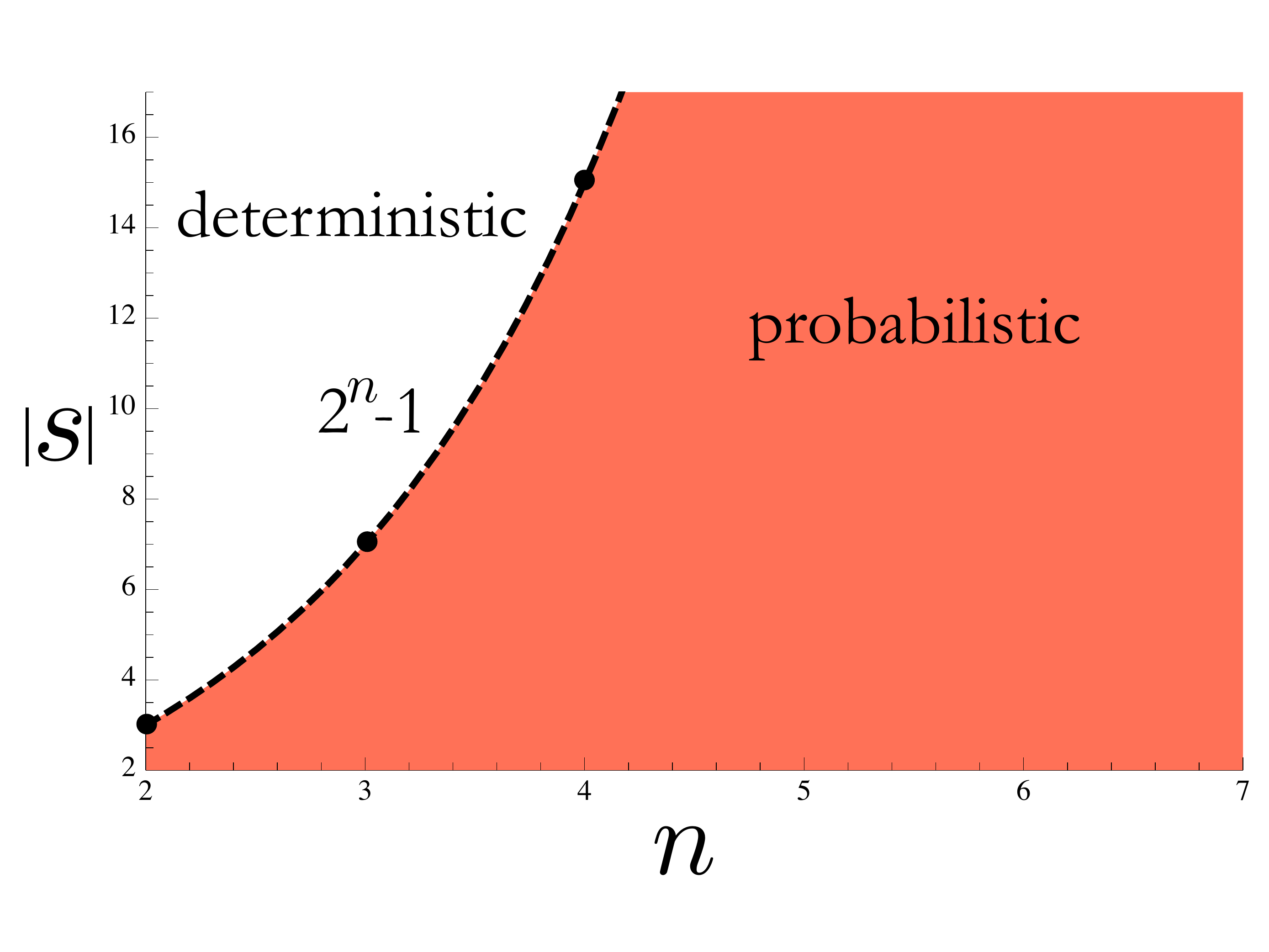} & \includegraphics[width=0.40\textwidth]{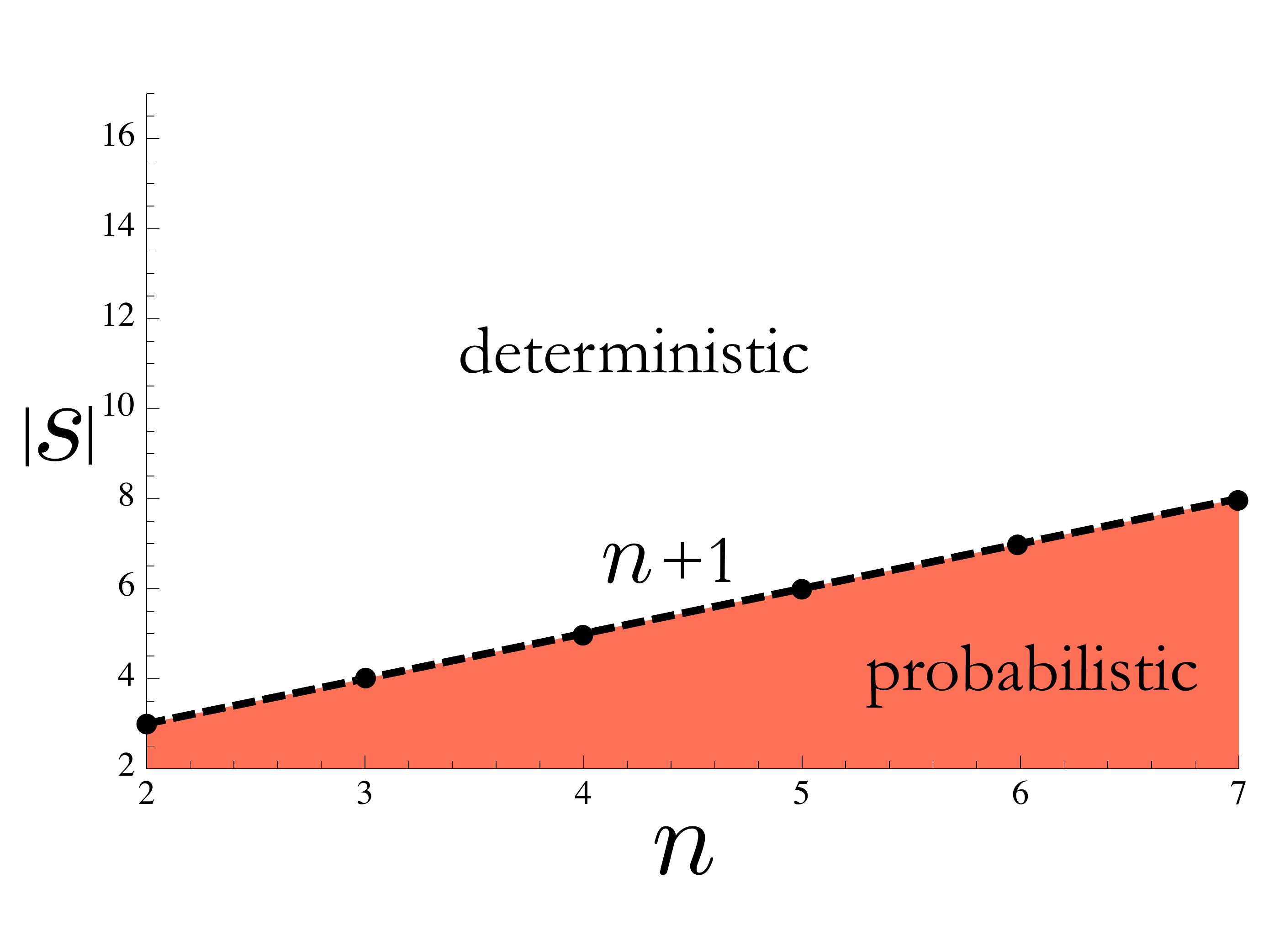} \\
\end{tabular}
\caption{Each graph compares the size in input $x$ to the available number of sites labeled by the size of $s$, the shaded regions indicate whether we can perform a particular function deterministically or probabilistically with quantum resources in NMQC$_{\oplus}$: a) for the $n$-tuple AND function we require the most resources ($(2^n-1)$ measurement sites) that are required for determinism therefore the region under this line is completely probabilistic; b) for the pairwise AND function we require exponentially fewer resources than for the $n$-tuple AND for deterministic computation. The regions are continuously shaded for clarity.}
\label{fig:graphs}
\end{figure}

Proposition 1 shows that the upper bound in resources given by Theorem 1 is tight for certain Boolean functions.  Proposition 2 on the other hand shows that we can perform certain functions deterministically with just a single ancilliary qubit. It is striking that the two families of functions have such divergent behaviour. We can see graphically in Figure. \ref{fig:graphs}, the contrast between the resources required for each function to produce deterministic and probabilistic computations.
\newline

\noindent
\textbf{Corollary 2:} Proposition 1 shows that certain families of Boolean functions are impossible to achieve efficiently and deterministically in {{\nmqc}}. This implies that it is impossible to efficiently achieve universal quantum computation deterministically in {{\nmqc}}. 
\newline

This verifies the long-held intuition that adaptive measurement is important for MQC. However, we note that this corollary only has relevance for MQC with linear side-processing. It tells us nothing of the full power of non-adaptive MQC with non-linear side-processing. This remains an open question.

The Popescu-Rohrlich non-local (PR) box is a widely studied \cite{pr, barrett} example of a correlation resource which is not achievable in quantum mechanics. In the language of {{\nmqc}} it is simple to define; it is the correlation which allows one to calculate the two-bit AND function $f(x)=x_1x_2$ via two measurements alone (such that marginal distributions are unbiased and hence non-signalling). Although the PR-box plays a central role in the study of bi-partite correlations, recently there has been much interest in the many-body correlations supported in theories more general than quantum mechanics \cite{barrett2}. The most general such framework, in which all non-signalling correlations are permitted, is commonly called ``boxworld'' or a Generalised Non-signalling Theory \cite{barrett2}. We can use the {\nmqc} framework to classify some of the correlations in such a theory. For example Propositions 1 and 2 immediately identify that any correlation which permits in {\nmqc} the computation of a function with fewer measurement sites than the lower bounds proven here, exists only in boxworld and not in quantum theory. The {\nmqc} framework may therefore be useful in the study and categorisation of some of the many-party generalisations of the PR-box.

\section{Non-adaptive MQC and Bell Inequalities}\label{nondet}

\noindent
We have seen that deterministic computations in {\nmqc} naturally define generalisations of the GHZ paradox. In this section, we shall turn our attention to non-deterministic computation, where we see a close relation with multi-party generalisations of the CHSH inequality. We will consider non-deterministic games of the following form. An input bit-string $x$ is supplied according to probability distribution $w(x)$. The goal of the game is to return the output of the Boolean function $f(x)$ within the framework of {\nmqc}. In this case, we do not demand that the game is won deterministically, instead we accept a less than unity success
probability, and wish that this probability is the highest possible. Specifying an upper bound to the mean success probability, in the case of a classical or LHV correlation, is equivalent to defining a set of Bell inequalites \cite{cleve}. Similarly, specifying an upper bound with quantum resources is equivalent to a generalisation of Tsirelson's bound \cite{tsirelson}. Note that for non-deterministic games it is critical that the prior probability distribution $w(x)$ over the inputs $x$ is specified, since this defines the mean success probability which is the figure of merit for the game. This family of games were first defined, and studied in depth, in \cite{cleve}.

In {\nmqc} every measurement has two-settings and two-outcomes. Multi-partite generalisations of the CHSH inequality for dichotomic two-setting measurements were defined and studied by Werner and Wolf \cite{WW} and independently by \.{Z}ukowski and Brukner \cite{ZB}. A traditional way to define Bell inequalities is as a bound on a linear combination of conditional expectation values $\epsilon(s)$, where $s$ represents the bit-string defining the measurement settings, and $\epsilon(s)$ is the expectation value of $(-1)^{\oplus_j m_j}$ with the outcome of the $j^{\textrm{th}}$ measurement as bit $m_j$. In other words, using $+1$, and $-1$ instead of 0 and 1 to denote measurement outcomes $\epsilon(s)$ represents the expectation value of the product of the outcomes. In our notation, $\epsilon(s)$ is defined in terms of probabilities $p(\oplus_j m_j|s)$:
\begin{equation}\label{epsilondef}
	\epsilon(s)=p(\oplus_j m_j=0|s)-p(\oplus_j m_j=1|s) .
\end{equation}

\noindent
The generalised CHSH inequalities of Werner, Wolf, \.{Z}ukowski and Brukner (WW\.{Z}B) take the form
\begin{equation}\label{WWZB}
\sum_{s}\beta(s)\epsilon(s)\leq u,
\end{equation}

\noindent
where $\beta(s)$ are real coefficients, and $u$ is a real value which is the upper bound to this quantity. When the correlation is due to a LHV source, we shall label the upper bound $u=c$. For the upper bound for measurements on a quantum state, we use the notation $u=q$. In this way, equation \eqref{WWZB} incorporates both Bell inequality and Tsirelson bound. The normalisation of $\beta(s)$ is arbitrary, and can be chosen as best suits the calculations in hand. Werner and Wolf normalise the values of $\beta(s)$ such that $c$ is always unity. To simplify our expressions here, however, we choose a different normalisation, and set $\sum_{s}|\beta(s)|=1$. This can be achieved for any CHSH-type inequality of form \eqref{WWZB} via a simple rescaling.

 Werner and Wolf in \cite{WW} give a compact expression which allows $q$ to be derived for a given set of coefficients $\beta(s)$ from a simple variational formula.
The connection between such Bell inequalities and {\nmqc} is straightforward and expressed via the theorem below. In fact, it suffices to consider a restricted form of {{\nmqc}}, which we shall label {{\nmqcs}}, in which pre-processing of the input data $x$ is fixed to the identity map (i.e. $s=x$). In discussing {{\nmqcs}}, we shall use $x$ and $s$ interchangably.
\newline

\noindent
\textbf{Theorem 2} Every multipartite Bell (Tsirelson) Inequality of the form of equation \eqref{WWZB} (normalised to $\sum_{s}|\beta(s)|=1$)  can be re-expressed as a bound on the maximum mean probability $\bar{p}$ of an {{\nmqcs}} computation of a Boolean function $f(x)=\textrm{sign}(\beta(s))$ with $s=x$ and $n=|s|$ measurements, where input bits $s$ are supplied according to probability distribution $w(s)=|\beta(s)|)$,
\begin{equation}
\bar{p}\le\frac{u+1}{2} ,
\end{equation}

\noindent
where this particularly simple form is due to the choice of normalisation $\sum_{s}|\beta(s)|=1$.
\newline

\noindent
\textit{Proof of Theorem 2} The proof of this theorem is similar in  content to observations made in \cite{vandam}, \cite{cleve} and \cite{brukner}, and, indeed, the well-known mapping of the CHSH inequality to the CHSH game. However the proof is simple, so we shall provide it here. 
We begin by substituting equation \eqref{epsilondef} into equation \eqref{WWZB}. Using the law-of-excluded middle for binary outcomes $p(\oplus_j m_j=0|s)+p(\oplus_j m_j=1|s)=1$ and some simply algebra, we can rewrite equation \eqref{epsilondef} as
\begin{equation}\label{epsilondef2}
	\epsilon(s)=(-1)^{f(s)} (2p(\oplus_j m_j=f(s)|s)-1)
\end{equation}
which is true for any Boolean function $f(s)$. Substiting this into equation \eqref{WWZB}, and letting $f(s)=\textrm{sign}(\beta(s))$, and $w(s)=|\beta(s)|/(\sum_s|\beta(s)|)$  we obtain,
\begin{equation}
|\sum_{s} w(s) p(\bigoplus_{j=1}^{m}\mu_{j}=f(s)|s)|\leq\frac{u+\sum_{s}|\beta(s)|}{2\sum_s |\beta(s)|} .
\end{equation}

\noindent
Werner and Wolf showed that the optimal quantum bound $u=q$ is always achieved by performing measurements on a GHZ state \cite{WW}, thus echoing the optimal quantum resource state in Theorem 1.

The expression on the left is the mean success probability of computing the function $f(s)$ in the {{\nmqcs}} model. Note that in this model, the only possible post-processing step for any function which does not act trivially on any input bits, is the XOR of all output bits, or the NOT of this bit. This completes the proof.$\square$
\newline

Theorem 2 informs us that the quantum game corresponding to any WW\.{Z}B Bell inequality is a computation in {\nmqc}. Or oppositely put, the ability for a computation to be achieved in {{\nmqcs}} can always be represented as a Bell inequality. Note also, that since pre-processing can be incorporated in the prior probability distribution $w(s)$, this result can be generalised, in a straightforward way, to the full class {\nmqc}. 

The {\nmqc} setting thus provides a frame-work in which we study families of Bell inequalities, over arbitrary numbers of parties, and calculate some of their generic properties. As an example of this, we shall return to the two families of functions defined in the previous section, the $n$-bit pairwise AND, and the $n$-tuple AND. Just as in the deterministic case these families provide interestingly contrasting behaviour. Note that both families include for the $n=2$ case, the bi-partite AND function, and hence embody the standard CHSH inequality. We shall limit our study to the case where inputs $s$ satisfy  a uniform distribution, i.e. where $w(s)=2^{-n}$.
\newline

\noindent
\textbf{Proposition 3:} The {\nmqcs} quantum game defined by $n$-tuple AND function $g_n(s)$ has {no quantum advantage} (i.e. $q=c$) for $n\geq 3$ if inputs are supplied from a uniform distribution $w(s)=2^{-n}$. For $n>2$ the upper bounds on mean success probability are $(1-\frac{1}{2^{n}})$ for both quantum and LHV cases.
\newline

This result is proved in section 2 of Appendix \ref{app_prop3}. This upper bound can be achieved by a computer which always outputs 0, so the quantum resources can be considered to provide no advantage for this computation whatsoever. Strikingly, the CHSH inequality (for $n=2$) is the \emph{only} member of this family where quantum mechanics exceeds the LHV bound.
\newline

\noindent
\textbf{Proposition 4:} The {\nmqcs} quantum game defined by the multipartite pairwise AND function $h_n(s)$ on $n$-bits supplied from a uniform distribution $w(s)=2^{-n}$
 possesses a {quantum advantage} with respect to LHV resources for every $n$. The ratio of the upper bounds for the associated Tsirelson/Bell inequalities $q/c$ grows exponentially in $n$. 
\newline

This can be seen by recognising that the Bell inequalities defined by $h_n(s)$ on a uniform distribution are equivalent (up to the relabelling of input and output bits) to the well-studied generalised Svetlichny inequalities \cite{collins2002}, which can be verified via some algebraic manipulation (a similar fact is noted here \cite{reznik}). It is well known that the ratio of quantum to classical upper bounds for these inequalities scales exponentially in $n$. In fact, when $n$ is even these inequalities are equivalent to the Mermin inequalities \cite{mermin} known \cite{WW} to possess the maximal ratio of quantum to classical bounds of any inequality of form \eqref{WWZB}. For even $n$, the bounds are $c=2^{\frac{-n}{2}}$ and $q=2^{\frac{-1}{2}}$. For odd $n$, $c=2^{\frac{1}{2}(1-n)}$ and $q=2^{\frac{-1}{2}}$.

\section{Discussion}\label{discussion}

\noindent
We have shown that there is a concrete connection between many-party Bell inequalities, and non-adaptive measurement-based quantum computation, and, indeed, the {\nmqc} model provides a concise method for defining Bell inequalities via quantum games. The framework we developed unified a number of known results, and led to several new results and insights, both with implications for measurement-based computation and Bell inequalities. The prominent appearance of the GHZ states in this work should not be a surprise. Werner and Wolf showed that \cite{WW} such states always provide the optimal quantum bounds for Bell inequalities of the form of equation \eqref{WWZB}. Theorem 2 thus implies that these states will always be optimal in \nmqc. The broader family of Reed-Muller states introduced by Raussendorf in \cite{raussendorf} are not required for these purposes.

There are a number of ways in which one might generalise these results. Firstly, it is natural to investigate whether higher-dimensional variations of MQC \cite{miyake} might share the same relationship with Bell inequalities for more than two settings and outcomes per measurement. Higher-dimensional MQC is a straight-forward generalisation of the qubit case. However, there are certain difficulties which will need to be overcome to make this connection. Firstly, while higher-dimensional GHZ paradoxes have been identified \cite{kasz}, in none of these cases do the correlations represent a full non-linear function, but correlations only hold for a subset of possible input strings. Whether this subset might be identified with pre-computation and whether or not higher-dimensional GHZ paradoxes might be found which achieve a full function in higher dimension {\nmqc} remains a question for future study. Secondly, by considering computations with multiple output bits, one could start to explore the full region of conditional output probabilities for both the quantum (i.e. generalise the Tsirelson-Landau-Masanes region \cite{tsirelson, masanes}) and LHV case.

Recently, attempts have been made to axiomatise quantum physics \cite{pr} via information theoretic approaches \cite{barrett}. Principles such as Information Causality \cite{pawlowski} and Macroscopic Locality \cite{navascues} have been successfully applied to derive, or at least bound, the set of correlations permitted under quantum mechanics in the bi-partite setting. These approaches, while very successful for the bi-partite case, do not easily generalise to the multi-party setting. Since the {\nmqc} framework naturally poses Bell inequalities in an informational theoretic framework, this could be a fruitful starting point for axiomatic reconstructions of multi-party quantum correlations.

A related idea, considered earlier by Van Dam\cite{vandam}, and Brassard et al \cite{brassard}, was to question whether the limitations on quantum correlations, expressed by Tsirelson-type bounds, might be imposed by Nature, in order to prevent distributed computation being achieved with trivial communication complexity. It was noted by Van Dam that if PR-boxes existed in Nature, then arbitrary distributed Boolean functions between two separated players, could be achieved with just a single bit of communication. Brassard et al argued that perhaps Tsirelson's bound on the CHSH inequality might take precisely that value which precludes such a dramatic collapse of communication complexity. They showed that, using Von Neumann's majority code error correction scheme, that there existed potential values of Tsirelson's bound less than 4 (but higher than $2\sqrt{2}$) at which communication complexity for all functions would be a single bit. However, this argument did not recover Tsirelson's bound exactly.

Recently, a multi-partite generalisation of this approach was considered by Marcovitch and Reznik \cite{reznik}. They showed that the Von Neumann majority gate correction scheme could be generalised to many-party functions by replacing the PR box by a non-local box defined by the pairwise AND function $h_n(x)$. They showed that, the assumption that such error correction be impossible, places an upper limit on the quantum bound for the generalized Svetlichny inequalities, providing evidence that Brassard et al's conjecture might hold in the multi-party case. Our results, have interesting implications for such arguments. Since for $n$-tuple AND functions quantum and classical bounds coincide (under the assumption, implicit in the above works, that the input distribution is uniform), any bounds derived via  communication complexity for quantum correlations must also be tight for classical correlations. If the communication complexity conjecture were true, this would imply that the classical upper bound for the distributed computation would place a lower bound on the (classical) threshold for Von Neumann's error correction scheme. In other words, two classical bounds would be causally related, but related via a quantum argument.

The approaches taken here are typical of recent attempts in quantum information science to use computer science formalisms to clarify, simplify and generalise results in physics. Thinking of Bell inequality experiments as computations has been very fruitful, and it seems likely that such approaches will continue to provide fresh insights into long-studied physics problems.

\textbf{Acknowledgements} We would like to thank Janet Anders for many useful discussions at the outset of this work, and for interesting discussions along the way we thank Miguel Navascu\'{e}s, Paul Skrzypczyk, Marcin Pawlowski, Mafalda Almeida and Joel Wallman. MJH acknowledges financial support from the EPSRC, ETC is supported by the Royal Commission for the Exhibition of 1851 and the EU integrated project Q-ESSENCE, KL is supported by the Materials Department, University of Oxford and St. Edmund Hall and DEB acknowledges financial support from the Leverhulme Trust and the National Research
Foundation and Ministry of Education, Singapore.

\appendix


\section{Equivalence of P matrices}\label{app_P_ Equivalence}

\noindent
The $P$ matrix (size $n$-by-$|s|$) specifies the linear dependence between the measurement settings, $s$, and the input, $x$.  This appendix demonstrates that many choices of $P$ are equivalent.  Whenever $|s|<2^n-1$, this equivalence allows all $P$ to be brought into a form where no column has all entries equal to one,  formally $\prod_{k} P_{j,k}=0, \forall j$. This property is exploited when we prove Proposition 1.  We begin by proving the following lemma.

\begin{lem}
A  {{\nmqc}} computation with $|s|$ measurements and with pre-processing map $P$ can deterministically compute a function $f(x)=\prod_{j=1}^{n}(x_{j}\oplus1)$ if and only if there exists a computation with pre-processing map $P'=(PM)_{\oplus}$ which can also deterministically compute $f(x)$, where $M$ is any invertible binary $n$-by-$n$ matrix.  Hence, the impossibility of a deterministic computation with $|s|$ measurements in \nmqc with pre-processing map $P$ implies the impossibility of computation with $|s|$ measurements by any  preprocessing map $P'=(PM)_{\oplus}$ .
\end{lem}

\noindent
Recall that a function is deterministically computed whenever \eqref{theorem1key} is satisfied, such that:
\begin{equation}\label{a1}
	e^{i   ( P x )_{\oplus}^{T} \phi}={-1}^{f(x)+c}, \forall x \in \{ 0, 1 \}^{n}.
\end{equation}
Taking the action of a binary, invertible matrix $M$ the above expression is altered to give
\begin{equation}
	e^{i   ( P Mx )_{\oplus}^{T} \phi}={-1}^{f(x)+c}, \forall x \in \{ 0, 1 \}^{n}.
\end{equation}
We notice that the set $\{x | x \in \{ 0, 1 \}^{n} \}$ is identical to $\{ Mx | x \in \{ 0, 1\}^{n} \} $ for any invertible binary $n$-by-$n$ matrix $M$. Secondly, $f(x)=f(Mx)$ for the function $f(x)=\prod_{j=1}^{n}(x_{j}\oplus1)$ since the $x=\{0,0,...,0\}$ all zeroes case is unaffected by these linear transformations $M$. Hence, if the equation \eqref{a1} is satisfied by $P$ and $\phi$, it is also satisfied if $P$ becomes $(PM)_{\oplus}$.

Next we show that (assuming  $|s|<2^n-1$) all $P$ are equivalent to some $P'=(PM)_{\oplus}$, such that the desirable property $\prod_{k} P'_{j,k}=0, \forall j$ holds.  Denoting the $j^{\mathrm{th}}$ row of $P$ by $[P]_{j}$, we observe that, since $|s|<2^n-1$, the rows $[P]_{j}$ will not include every possible non-trivial binary string.  We label one of the missing binary strings as $w$.  There always exist an invertible $M$ such that $wM= \{1,1,...1\}$.  Since the map is invertible all rows $[P]_{j} \neq w$ are mapped to $[PM]_{j} \neq \{1,1,...1\}$.

\section{Proof  of Proposition 1}\label{app1}
\textbf{Proposition 1:} To achieve the $n$-tuple AND function $g_{n}(x)$ deterministically in  {{\nmqc}}  one requires no fewer than $(2^{n}-1)$ qubits.
\newline

For the purposes of the proof it is simpler to consider the function $k_{n}(x)=1\oplus\prod_{j}^{n}(x_{j}\oplus1)$. This function is obtained from $g_{n}(x)$ under local relabelling (NOT) of the input bits. Thus bounds we prove for $k_{n}(x)$ will thus also hold for $g_{n}(x)$.  For deterministic evaluation of the function, equation \eqref{theorem1key} must be satisfied, which we assume to be true and then derive a contradiction.

Assuming that equation $\eqref{theorem1key} $ holds for $x=\{ 0,0,..0\}$ entails
\begin{equation}
	e^{0}=(-1)^{k_{n}(0,0...0)+c}=(-1)^{c},
\end{equation}
and hence $c=0$.  Next, assuming  $\eqref{theorem1key} $ holds for $x \neq \{ 0,0,..0\}$ (for which $k_{n}(x)=1$) entails
\begin{equation}
	e^{i (Px)_{\oplus}^{T}\phi }=(-1)^{k_{n}(x)}=(-1), \forall x \neq{0,0,..0},
\end{equation}
and hence
\begin{equation}
(Px)_{\oplus}^{T} \phi =  \pi (2 t_{x} +1), \forall x \neq{0,0,...,0},
\end{equation}
for some integer $t_{x}$.  We can construct a sum over all $x$ which alternates in sign:
\begin{equation}\label{earl2}
 \left[	\sum_{x \in \{0,1\}^{n}} (-1)^{W(x)}(Px)_{\oplus}^{T} \right]\phi =  \sum_{x|x \neq 0} (-1)^{W(x)} [ \pi(1 + 2 t_{x} ) ] 
 \end{equation}
Collecting terms on the R.H.S. which have the Hamming weight, $W(x)$, equal to $y$, and defining $t'_{y}=\sum_{\{ x ; W(x)=y \}}  t_{x}$, gives a summation over numbers
\begin{eqnarray}
	& = & \sum_{y = 1}^{n} (-1)^{y}  \pi \left( \frac{n!}{y!(n-y)!} + 2 t'_{y} \right)   , \\
	& = &  \pi(2t - 1) ,
\end{eqnarray}
where $t$ is again an integer, such that $t = \sum_{y=1}^{n} (-1)^{y} t'_{y}$.  Our proof proceeds by showing that the L.H.S. of \eqref{earl2} vanishes, and therefore results in the contradiction that $\pi(2t - 1)=0$.  The $j^{\mathrm{th}}$ element of the L.H.S. of \eqref{earl2} is
\begin{equation}
\label{eqn_sumP}
	 \left[	\sum_{x \in \{0,1\}^{n}} (-1)^{W(x)}(Px)_{\oplus} \right]_{j} =  \sum_{x \in \{0,1\}^{n}} (-1)^{W(x)} ( \bigoplus_{k} P_{j,k}x_{k} ) .
\end{equation}
Next we observe that this vanishes if for any $k$,  we have $P_{j,k}=0$.  To see this we assume without loss of generality that $P_{j,n}=0$.  Consider pairs of terms for $x'$ and $x''$ which are identical except for the $n^{\mathrm{th}}$ element, such that $x'=\{x_{1}, x_{2},... x_{n-1},0 \} $ and $x''=\{x_{1}, x_{2},... x_{n-1}, 1 \} $.  For these terms $( \bigoplus_{k} P_{j,k}x'_{k} )  = ( \bigoplus_{k} P_{j,k}x''_{k} )  $ since $P_{j,n}=0$ the value of the $n^{\mathrm{th}}$ element does not matter.  However, $W(x'')=W(x')+1$, and so $(-1)^{W(x')}=-(-1)^{W(x'')}$ and so the two terms cancel.  The whole summation can be broken up into distinct pairs of the form, $x'$ and $x''$, and since each pair cancel, the whole sum must vanish.

Recall that for our proof by contradiction we wish to show the vector of equation \eqref{earl2} is zero for every element (every $j$).   The preceding argument shows that the $j^{\mathrm{th}}$ element will vanish unless $\prod_{k}P_{j,k}=1$; that is, the $j^{\mathrm{th}}$ row of $P$ is a row of all ones.  However, Appendix~\ref{app_P_ Equivalence} demonstrated that every $P$ is computationally equivalent to many other $P'$ matrices.  Furthermore, whenever $|s|<2^{n}-1$, we can find an equivalent $P'$ for which no row contains all ones.   This completes the proof. $\square$
\newline

To see why the proof fails for $(2^n-1)$ distinct measurements, we note that in this case, $P$ must always have the all-ones row (under the assumption that all rows are distinct and no row contains all zeros), and then a solution to equation \eqref{theorem1key} may always be found, as shown explicitly in the proof of Theorem 1.


\section{Proof  of Proposition 3}\label{app_prop3}

\noindent
\textbf{Proposition 3:} The {\nmqcs} quantum game defined by $n$-tuple AND function $g_n(s)$ has {no quantum advantage} (i.e. $q=c$) for $n\geq 3$ if inputs are supplied from a uniform distribution $w(s)=1/(2^n)$. For $n>2$ the upper bounds on mean success probability are $(1-\frac{1}{2^{n}})$ for both quantum and LHV cases.
\newline

\noindent
To prove this, we shall explicitly derive the Tsirelson bound for this particular family of Bell inequalities and show that it is equal to the bound obtained by the trivial classical strategy, where every measurement  device always outputs zero. In WW\.{Z}B form, with normalisation $\sum |\beta(s)|=1$, the inequality may be (according to Theorem 2) written
\begin{equation}
|\sum_{s}\frac{(-1)^{\prod_{k=1}^{n}s_{k}}}{2^{n}}\epsilon(s)|\le c\le q.
\end{equation}
For the trivial classical strategy, $\epsilon(s)=1$ for all $s$, and thus quantity on the left hand side of the inequality is equal to $(2^{n}-2)/2^{n}$. To prove the proposition, we will show that the quantum upper bound (and hence also the classical upper bound) for this quantity is also equal to $(2^{n}-2)/2^{n}$. Werner and Wolf show in \cite{WW} that $q$ must satisfy the following equation
\begin{equation}\label{wwbound}
q= \underset{{\{\phi_{k}\}}}{\textrm{sup}} |\sum_{s}\frac{(-1)^{\prod_{k=1}^{n}s_{k}}}{2^{n}}\prod _{k=1}^{n} e^{i s_{k}\phi_{k} }|.
\end{equation}
Due to the form of this equation, we may, without loss of generality, restrict $\phi_k$ to the range $\phi_k\in(-\pi,\pi)$. We simplify equation \eqref{wwbound}, using the fact that $(-1)^{\prod_{k=1}^{n}s_{k}}=+1$ for all bit strings $s$ except when $s$ is the ``all-ones'' bit-string, to write
\begin{equation}
2^n q= \underset{{\{\phi_{k}\}}}{\textrm{sup}} |\sum_{s}\prod _{k=1}^{n} e^{i s_{k}\phi_{k} }-2e^{i\sum_{k}^{n}{\phi_{k}}}|.
\end{equation}
The sum over $s$ then factorises to achieve
\begin{equation}\label{CHSHprobb}
2^n q=\underset{{\{\phi_{k}\}}}{\textrm{sup}} |2^{n}\prod_{k=1}^{n}\cos\left(\frac{\phi_{k}}{2}\right)-2e^{i\sum_{k}^{n}\frac{\phi_{k}}{2}}|
\end{equation}

\begin{figure}[htbp]
	\centering
		\includegraphics[height=3in]{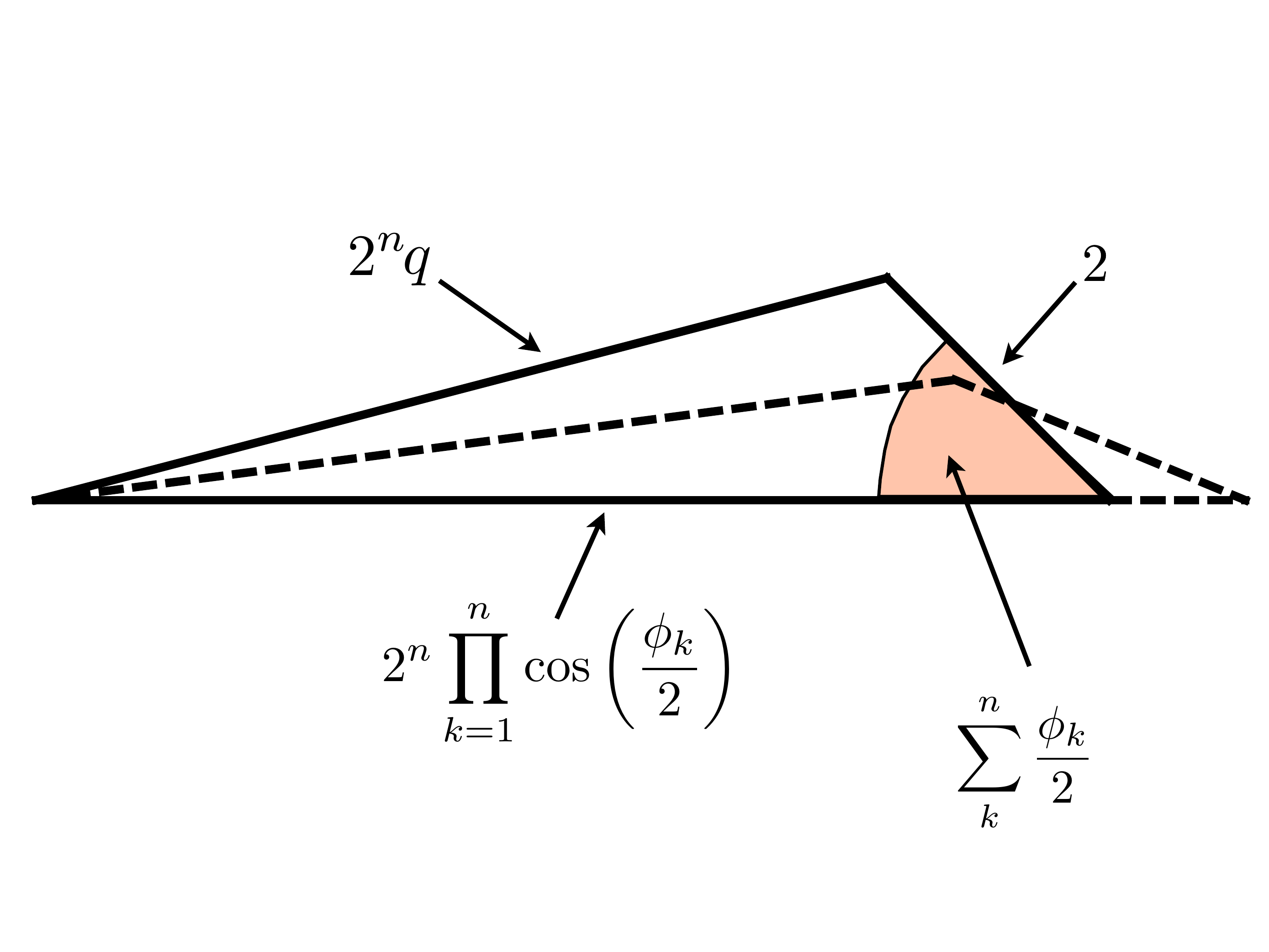}
	\caption{The quantity $2^n q$ in equation \eqref{CHSHprobb} represented geometrically as the length of the third side of this triangle. A simple geometric argument in the text shows that when $n>7$, this side cannot exceed $2^n-2$.}
	\label{fig:triangle}
\end{figure}

We now adopt a geometric argument. The goal is to maximize the modulus of a sum of two complex numbers. These numbers may be represented, on the plane, as two sides of a triangle, as illustrated in figure~\ref{fig:triangle}. The first side has length $2^{n}\prod_{k=1}^{n}\cos(\frac{\phi_{k}}{2})$, the second is of length 2 and the angle between these sides is $\sum_{k}^{n}\frac{\phi_{k}}{2}$.  We complete the proof by showing that when $n>7$, the length of the third side of the triangle can never exceed $2^n-2$, and hence $q$ cannot exceed the classical value.

We proceed by assuming the opposite of what we want to prove and demonstrating a contradiction. For there to be a quantum violation of the classical bound we need $2^n q>2^n-2$ to be satisfied. Via the triangle inequality, for this inequality to be satisfied, the length of the base of the triangle must be greater than $2^n-4$, and thus
\begin{equation}\label{inequality}
	\prod_{k=1}^{n}\cos\left(\frac{\phi_{k}}{2}\right)>1-2^{2-n}.
\end{equation}

Since $\phi_k\in(-\pi,\pi)$ all terms in the product are non-negative, hence we can impose the weaker condition for \eqref{inequality} that $\forall\phi_{k}$,
\begin{equation}\label{ineq}
\cos\left(\frac{\phi_{k}}{2}\right)>1-2^{2-n}.
\end{equation}
 
which implies
\begin{equation}\label{ineqb}
-n\arccos(1-2^{2-n})< \sum_{k=1}^{n}\frac{\phi_{k}}{2}<n \arccos(1-2^{2-n})
\end{equation}
or
\begin{equation}\label{ineqcc}
|\sum_{k=1}^{n}\frac{\phi_{k}}{2}|<n \arccos(1-2^{2-n})
\end{equation}

Proceeding geometrically, we now use the cosine rule to find the length of the third side of the triangle
\begin{equation}\label{cosiner}
(2^nq)^2=4+2^{2n}\prod_{k=1}^{n}\cos^{2}\left(\frac{\phi_{k}}{2}\right)-2^{n+2}\cos\left(\sum_{l=1}^{n}\frac{\phi_{l}}{2}\right)\prod_{m=1}^{n}\cos\left(\frac{\phi_{m}}{2}\right).
\end{equation}

To obtain a quantum violation we need $(2^nq)^2>4+2^2n-2^{n+2}$. Since $\phi_k\in(-\pi,\pi)$ $\prod_{m=1}^{n}\cos(\frac{\phi_{m}}{2})$ is non-negative, and hence a violation can only be achieved if $\cos(\sum_{k=1}^{n}\frac{\phi_{k}}{2})$ is negative which implies  $|\sum_{k=1}^{n}\frac{\phi_{k}}{2}|>\frac{\pi}{2}$.

Using this, together with inequality \eqref{ineqcc}, we achieve
\begin{equation}
	n \arccos(1-2^{2-n})>\pi/2
\end{equation}
or equivalently
\begin{equation}\label{ineqc}
\cos\left(\frac{\pi}{2n}\right)>(1-2^{2-n}).
\end{equation}

\noindent
This inequality is only satisfied for integers $n\le 7$, hence, due to the contradiction with our initial assumption, for $n>7$ the quantum and classical bounds of the Bell inequality must coincide with that of the trivial strategy $(2^n-2)/2^n$, and there is no quantum advantage. Direct numerical verification of the bounds, via equation \eqref{CHSHprobb} for $n<7$ indicates that the bounds coincide for all integer values $3\le n\le 7$, completing the proof. $\square$
\newline

\end{document}